\documentclass[sigconf]{acmart}

\newcommand{\para}[1]{\smallskip\noindent\textbf{#1}}

\usepackage{lipsum}  
\usepackage{xcolor}
\usepackage{cleveref}
\usepackage{caption,subcaption}
\usepackage{multirow}
\usepackage{booktabs} 

\copyrightyear{2018}
\acmYear{2018} 
\setcopyright{iw3c2w3}
\acmConference[WWW '18 Companion]{The 2018 Web Conference Companion}{April 23--27, 2018}{Lyon, France}
\acmBooktitle{WWW '18 Companion: The 2018 Web Conference Companion, April 23--27, 2018, Lyon, France}
\acmPrice{}
\acmDOI{10.1145/3184558.3191567}
\acmISBN{978-1-4503-5640-4/18/04}

\begin{document}

\title{Towards Quantifying Sampling Bias in Network Inference}\titlenote{Please cite the WWW'18 version of this paper.}

\author{Lisette Esp\'in-Noboa}
\authornote{This work was part of her summer internship at USC-ISI in 2017.}
\affiliation{
  \institution{GESIS \& University of Koblenz-Landau}
}
\email{Lisette.Espin@gesis.org}
\author{Claudia Wagner}
\affiliation{
  \institution{GESIS \& University of Koblenz-Landau}
}
\email{Claudia.Wagner@gesis.org}
\author{Fariba Karimi}
\affiliation{
  \institution{GESIS \& University of Koblenz-Landau}
}
\email{Fariba.Karimi@gesis.org}
\author{Kristina Lerman}
\affiliation{
  \institution{USC Information Sciences Institute}
}
\email{lerman@isi.edu}

\renewcommand{\shortauthors}{Esp\'in-Noboa et al.}

\begin{abstract}

Relational inference leverages relationships between entities and links in a network  to infer information about the network from a small sample. This method is often used when global information about the network is not available or difficult to obtain. However, how  reliable  is   inference from a small labelled sample?  How should the network be sampled, and what effect does it have on inference error?  How does the structure of the network impact the sampling strategy? We address these questions by systematically examining how network sampling strategy and sample size affect accuracy of relational inference in  networks. To this end, we generate a family of synthetic networks where nodes have a binary attribute and a tunable level of homophily. As expected, we find that in heterophilic networks, we can obtain good accuracy when only small samples of the network are initially labelled, regardless of the sampling strategy. Surprisingly, this is not the case for homophilic networks, and sampling strategies that work well in heterophilic networks lead to large inference errors. 
These findings suggest that the impact of network structure on  relational classification is more complex than previously thought.
\end{abstract}



\maketitle

\section{Introduction}

Networks form the infrastructure of modern life, linking billions of people, organizations and devices via trillions of transactions. Solving today's problems and making critical decisions increasingly calls for mining massive data residing in such networks. Due to their size and complexity, it is often prohibitively costly for analysts to obtain a global view of the network and the data it contains. Instead, they can use relational machine learning methods to infer information about the network from a partial sample, e.g., infer the class of unlabelled nodes from the known classes of a few seed nodes. How reliable is such inference? How much impact does the choice of seeds have on inference error? How much does the structure of the network impact sampling strategy? In this work, we address some of these questions by systematically studying potential sources of bias in the relational inference process: Network Structure, Sampling, Relational Classification, and Collective Inference. Towards that goal, we dig deeper into how sampling strategies can be biased, and when this bias can be beneficial or disadvantageous for the inference process. New insights on how sampling impacts relational classification performance can potentially lead to new unbiased strategies.

\para{Relational Classification}. 
Relational classifiers  propagate information through the network from the known labels of nodes to infer unknown labels. The classification performance is measured by how well all the nodes' labels can be recovered when only the labels of a few seed nodes are known. 
In \cite{Macskassy2007} the authors outline the two main components of collective classification, which are the collective inference method and the relational classifier. They assess how various choices and combinations of components, as well as the percentage of labelled data used for training the method, impact the accuracy of the classification. Their results show that there are two sets of techniques that are preferable in different situations, namely when few versus many labels are known initially, and that link selection plays an important role similar to traditional feature selection. However, the authors did not  explore how the network structure impacts performance of collective classification methods.
Sen et al. close this gap by comparing four collective classification algorithms with a content-only classifier, which does not take the network into account, on networks that varied in link density and homophily \cite{sen08}. They found that increasing link density
improves the performance of collective classification and clearly outperforms content-only classifiers at all density levels. Moreover, \emph{homophily}, which refers to the tendency of nodes with similar labels to be connected, further helps collective classifiers outperform content-only classifiers, except for  very low levels of homophily ($<0.1$), where content-only classifiers perform slightly better.
While this work explores homophily and density of networks separately, more recent research investigates how these characteristics jointly impact performance. 
In \cite{zenoinvestigating} the authors show that as 
homophily and link density of the network increase, the accuracy of relational classification also increases. 
Similar to our study, that work focuses on balanced networks where nodes have a single binary attribute. However, the subgraph used for training is selected via random node sampling only.

\para{Sampling Bias}. Previous work has demonstrated that the estimates obtained from network samples collected by various crawlers can be inaccurate with respect to global \cite{leskovec2006sampling} and local network statistics \cite{Wagner2017}. Two recent papers showed that the choice of the initial sample of labelled seed nodes can also affect attribute inference \cite{ahmed2012network, Yang2017}. However, these did not explore how properties of networks, such as homophily, affect the choice of seeds and classification performance. 

\para{Findings and Contributions}. In this work, we focus on the attribute inference task and explore how the accuracy of collective inference in networks depends on the strategy used to create the initial set of labelled nodes. In summary, our main contributions are three-fold: (i) Using synthetic and empirical networks, we provide evidence that homophily plays a decisive role in the collective inference process: First, no sampling technique can beat a random classifier when networks are neutral (i.e., nodes connected at random). Second, heterophilic networks are easy to classify with any sampling strategy and require a training sample of at least 5\% of random nodes to achieve an unbiased classification. Finally, some sampling strategies that work well for heterophilic networks require larger samples for homophilic networks. Only methods that construct samples by selecting highest degree nodes first achieve good classification performance with small samples in both homophilic and heterophilic regimes.
(ii) We show that link density influences classification performance under certain conditions: First, sampling methods that rank low-degree nodes first, benefit from networks with high density. Second, high link density homophilic networks require larger training samples for edge, mixed degrees, and snowball sampling.
(iii) We discuss 
the impact of 
sampling strategies on relational classification using collective inference, and demonstrate that inference can be negatively affected by class imbalance.

The remainder of this paper is organized as follows: In \Cref{background} we present background knowledge. Experiments, datasets and results are described in \Cref{experiments} followed by a discussion in \Cref{discussion}. Finally, we present future work and conclusions in \Cref{conclusions}.

\section{Background}
\label{background}
In this work, we are primarily interested in studying the influence of different sampling techniques on relational classification. We describe (i) networks of interest, (ii) the classification process and (iii) used network sampling techniques.

\begin{figure}[tp!]
    \centering
    \begin{subfigure}[t]{0.2\textwidth}
        \centering
        \includegraphics[scale=0.3]{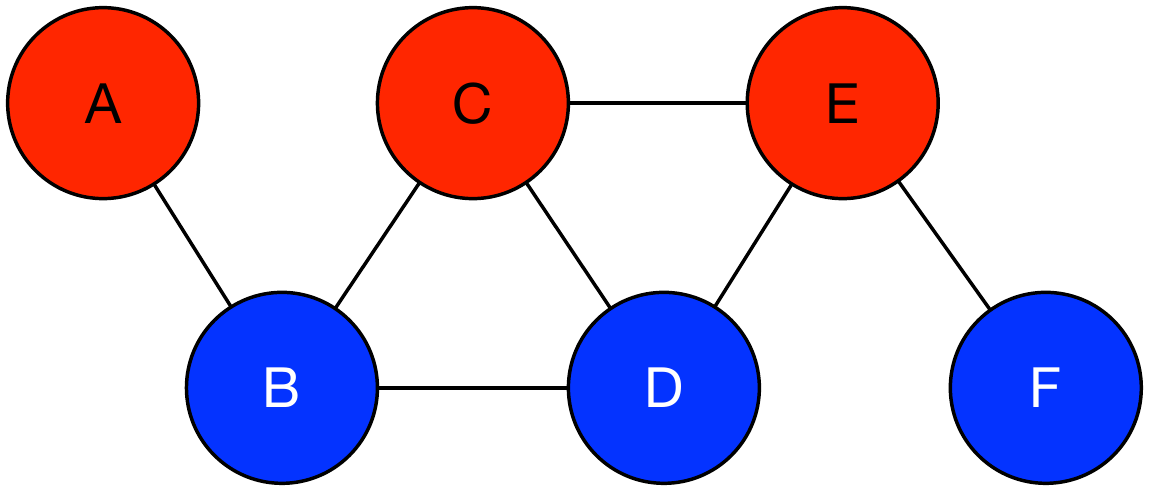}
        \caption{Network}
        \label{toy-example:all}
    \end{subfigure}
    \quad
    \begin{subfigure}[t]{0.2\textwidth}
        \centering
        \includegraphics[scale=0.3]{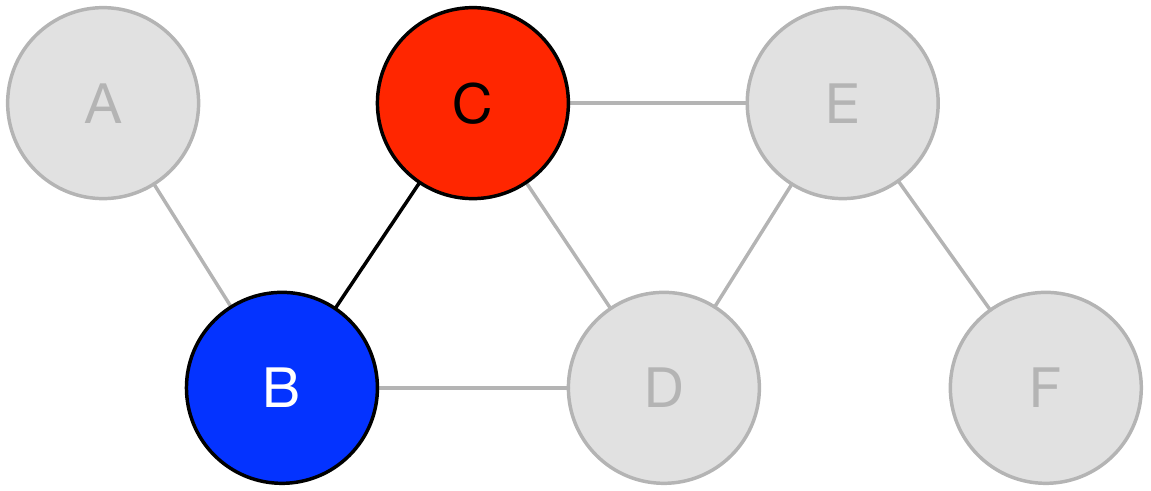}
        \caption{Sample \#1}
        \label{toy-example:sample1}
    \end{subfigure}
    \quad
    \begin{subfigure}[t]{0.2\textwidth}
        \centering
        \includegraphics[scale=0.3]{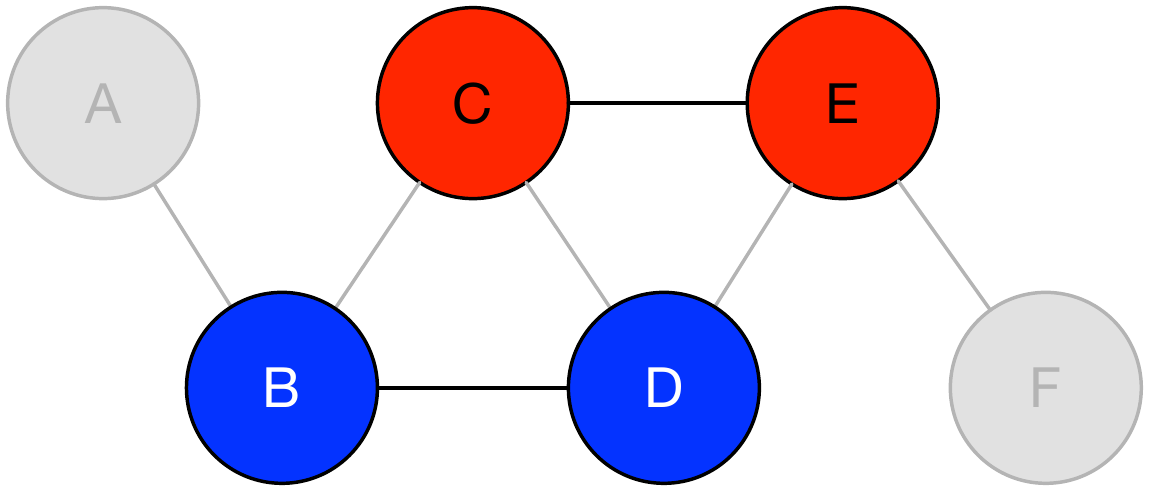}
        \caption{Sample \#2}
        \label{toy-example:sample2}
    \end{subfigure}
    \quad
    \begin{subfigure}[t]{0.2\textwidth}
        \centering
        \includegraphics[scale=0.3]{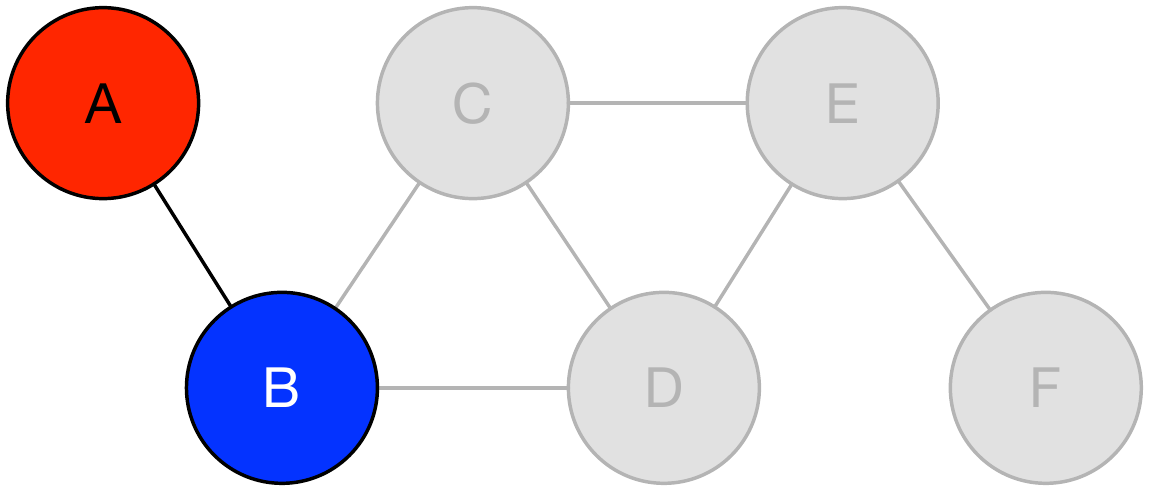}
        \caption{Sample \#3}
        \label{toy-example:sample3}
    \end{subfigure}
    \caption{\textbf{Example}. This figure illustrates an unweighted undirected node-attributed network and three different samples. (a) Shows a heterophilic network with seven edges and six nodes. Each node is coloured either red (A, C, E) or blue (B, D, F). (b) Sample \#1 shows a subgraph extracted by sampling two nodes. This sample includes nodes B and C, which reflect perfect heterophily. (c) Sample \#2 shows a homophilic subgraph sampled by randomly picking two edges, C-E and B-D. (d) This sample is similar to Sample \#1 as it reflects perfect heterophily. 
    However in this case node F is 3-HOPs away from the nearest seed node B (compared to 2-HOPs in Sample \#1).}
    \label{toy-example}
\end{figure}

\subsection{Attributed Networks}
We formally define this as: Let $G=(V,E,F)$ be an attributed unweighted graph with $V=(v_1,...,v_n)$ being a set of nodes, $E=\{(v_i,v_j)\} \in (V \times V)$ a set of either directed or undirected edges, and a set of feature vectors $F=(f_1,...,f_n)$. Each feature vector $f_i=(f_i[1],...,f_i[t])^T$ maps a node $v_i$ to $t$ (binary, numeric or categorical) attributes. A class label is defined as $c \in t$, and represents the attribute to be inferred in the classification process. Link density is described as the fraction of potential connections in G that are actual connections, that is $d=\frac{|E|}{N(N-1)}$ for directed, and $d=\frac{2|E|}{N(N-1)}$ for undirected networks. The average degree of the network captures the average number of edges per node: $\langle k \rangle=\frac{|E|}{N}$ for directed and $\langle k \rangle=\frac{2|E|}{N}$ for undirected networks. The network homophily $H$ is captured by the fraction of same-class connections among the total number of edges $|E|$. Homophily values range from $H=0.0$ to $H=1.0$. Networks with homophily $H=0.5$ are referred to as \emph{neutral}, otherwise they are \emph{heterophilic} if $H<0.5$, or \emph{homophilic} when $H>0.5$. Class balance $B$ captures the fraction of nodes under each class value. A network is \emph{balanced} when all class values have the same number of nodes, otherwise it is an \emph{unbalanced} network. 

In this work, we focus on attributed networks whose edges are unweighted and undirected, and whose nodes are balanced along a binary feature (e.g., $c = color \in \{blue, red\}$). \Cref{toy-example:all} shows an example of such network, where nodes are assigned to one color, either blue or red, and since only 2 out of 7 edges are same-color connections this network is heterophilic ($H\approx0.3$). This network is also balanced ($B=\frac{3}{6}=0.5$) since the number of blue nodes ($n_{b}=3$) is equal to the number of red nodes ($n_{r}=3$). 

Notice that in a realistic scenario, class balance $B$ is often unknown, as well as homophily $H$. However, these two values can be inferred. For instance, balance can be approximated by randomly picking a set of nodes to extract their true class value and then infer class balance. Similarly, homophily can be approximated by randomly picking a set of edges. These approximations go beyond the scope of this work. For evaluation purposes we assume total knowledge of the network. 

\subsection{Relational Classification}
Classification in networked data~\cite{Macskassy2007,getoor2007introduction,sen08} learns correlations between attribute values of linked nodes from observed data and then uses them in a collective inference process that propagates predictions through the network. This process can be divided into four phases. First, a subgraph needs to be \emph{sampled} from the network. Second, a \emph{local model} is learned by using information from nodes only (e.g., nodal attributes) and can be used as class priors later in the inference. Third, the \emph{relational model} in contrast to the local model, learns information from nodes and their 1-HOP neighbourhood. Finally, once the models' parameters (i.e., probabilities) have been learnt, the collective inference phase determines how the unknown values are estimated. 

Every phase can be implemented in different ways~\cite{Macskassy2007}. Since our work focuses on the sampling phase, we keep fixed the other modules by (i) learning the local model as class priors from the nodes in the training sample, (ii) learning the relational model from the nodes and edges in the training sample using Bayesian statistics, and (iii) inferring estimate values using Relaxation labelling. 

For simplicity, we focus on univariate network classification, which means that the structure of class linkage in
the network is modelled with no additional information from other attributes. This setup is referred to as network-only Bayes classifier (NBC) in \cite{Macskassy2007} to emphasize that local attributes of a node are ignored.

\subsection{Network Sampling}
\label{section:sampling}

The goal of sampling is to split the network into a \emph{training} and a \emph{testing} sample. First, a subgraph $\hat{G}=(\hat{V},\hat{E},\hat{F})$ is extracted from the network $G$ in order to learn the model parameters. Nodes $\hat{V} \subset V$ 
that belong to the training sample $\hat{G}$ 
are called \emph{seed nodes}, and we assume that their edges $\hat{E}=\{(\hat{v_i},\hat{v_j})\} \in (\hat{V} \times \hat{V}) \subset (V \times V)$  and attributes $\hat{F} \subset F$ are known by the classification algorithm. For example, based on the information shown in \Cref{toy-example}, if we choose the sample in \Cref{toy-example:sample1}, node A would be correctly classified as red, due to the fact that A is connected to a blue seed node, and the sample (C-B) reflects perfect heterophily. However, if we choose the sample in \Cref{toy-example:sample2}, node A would be classified as blue, because it is connected to a blue seed node and the sample (C-E, B-D) reflects perfect homophily. A different sample is shown in \Cref{toy-example:sample3}, in this case nodes A and B are selected as seed nodes, and regardless of the learnt model parameters (i.e., probability of connecting blue-blue, blue-red, red-blue, red-red), notice that node F is not connected to any seed node. Thus, the inferred attribute of node F will depend on the inferred attribute of node E, which in turn also depends on the estimates of unlabelled nodes, C and D. If those estimates are wrong the inference for node F will probably also be wrong.

Notice the importance of the sampling method. The selected nodes should not only reflect the global properties of the network such as balance and homophily but should also be as close as possible to the unlabelled nodes to avoid long label propagation chains that may potentially be erroneous.
Next, we describe ten different sampling methods that we evaluate in this work.

\para{Random Nodes}.
This is the most basic sampling method where a random fraction $p$ of nodes is selected. The sampled network then contains the selected nodes and all edges among them.

\para{Random Edges}.
This technique randomly selects edges from the set of all edges $E$. In order to make a fair comparison among other sampling techniques (based on number of nodes), we select edges randomly until we reach a specific fraction $p$ of nodes. That is why this sampling method is referred to as nedges.

\para{SnowBall}.
Snowball sampling~\cite{goodman1961snowball,atkinson2001accessing} randomly selects a starting node and all its neighbours as well as their neighbours' neighbours (similar to breadth-search-first). The algorithm continues until it has gathered a fraction $p$ of nodes.

\para{Degree}.
We rank all nodes by their degree in descendant (degreeDESC) and ascendant (degreeASC) order. The idea is to verify whether high (or low) degree nodes are good seeds for classification. Therefore, the fraction $p$ of selected nodes includes the top $p\times 100\%$ of nodes in the ranking. We also provide a mix of degrees (degreeMIX) by selecting $\frac{p}{2}\times 100\%$ of both top high and top low degree nodes.

\para{PageRank}. 
Similar to sampling by degree, we rank nodes by their PageRank (PR)~\cite{page1999pagerank} in descendant (pagerankDESC) and ascendant (pagerankASC) order. By using highest PR first, (pagerankDESC) we test whether the most important nodes in the network are good samples for the learning and testing inference. We expect pagerankASC to work poorly since their top low PR nodes are not well connected, and often have low degree.

\para{Optimal Percolation}.
The motivation behind optimal percolation~\cite{morone2015influence}, is to find a minimal set of nodes, called influencers, which, if activacted, would cause the spread of information through the whole network. Therefore, we rank nodes based on their \emph{collective influence} in descendant (percolationDESC) and ascendant (percolationASC) order. 

\begin{figure*}[hpt]
    \centering
    \includegraphics[height=0.45in]{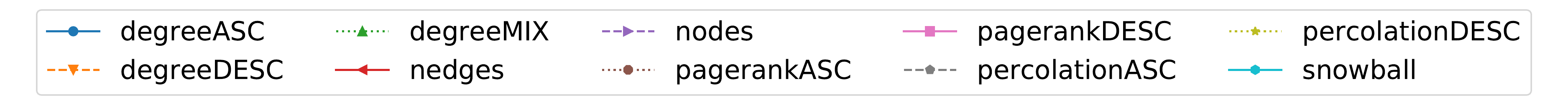}
    
    \centering
    \begin{subfigure}[t]{0.32\textwidth}
        \centering
        \includegraphics[height=1.8in]{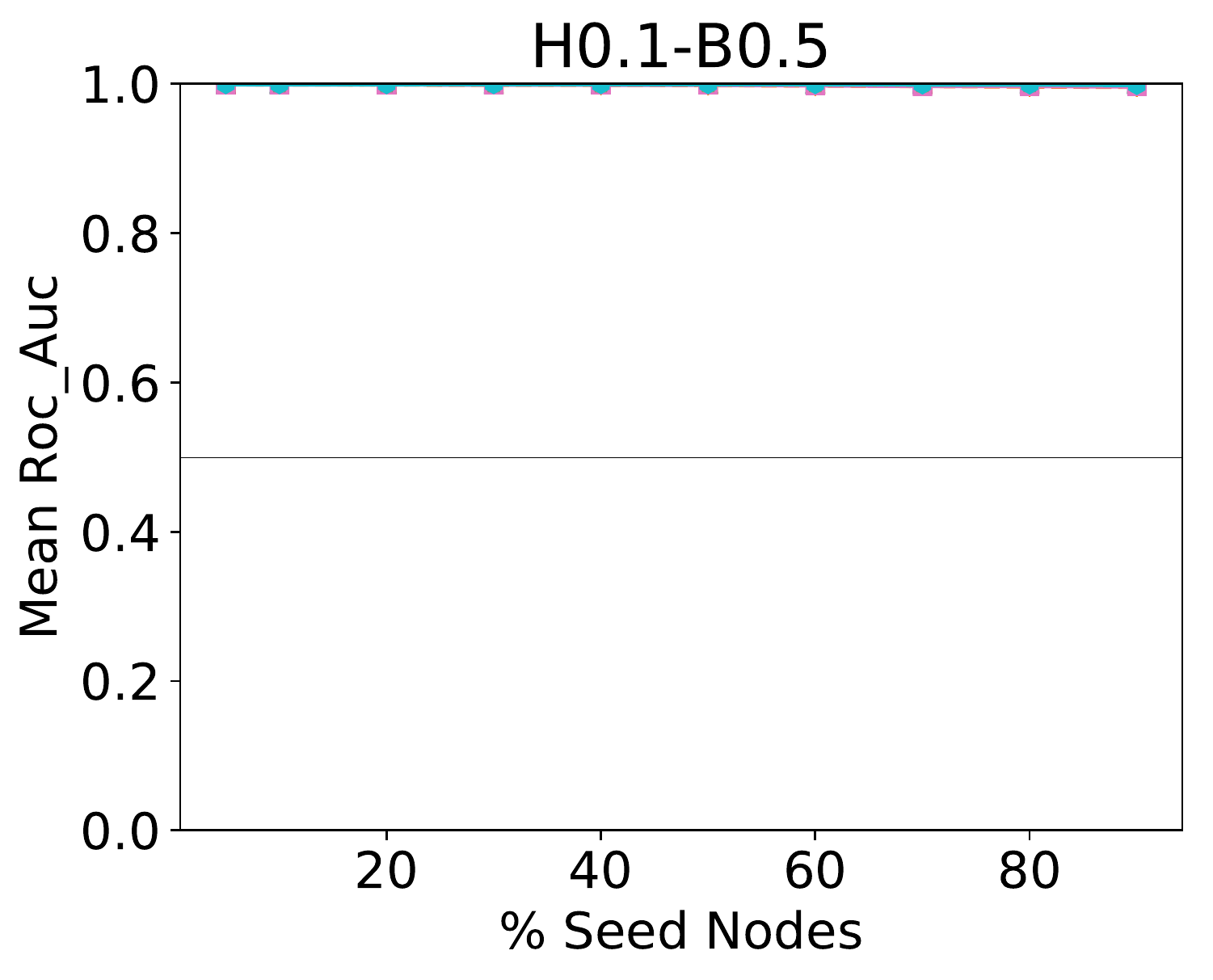}
        \caption{Heterophilic}
        \label{synthetic-rocauc:hete}
    \end{subfigure}
    ~ 
    \begin{subfigure}[t]{0.32\textwidth}
        \centering
        \includegraphics[height=1.8in]{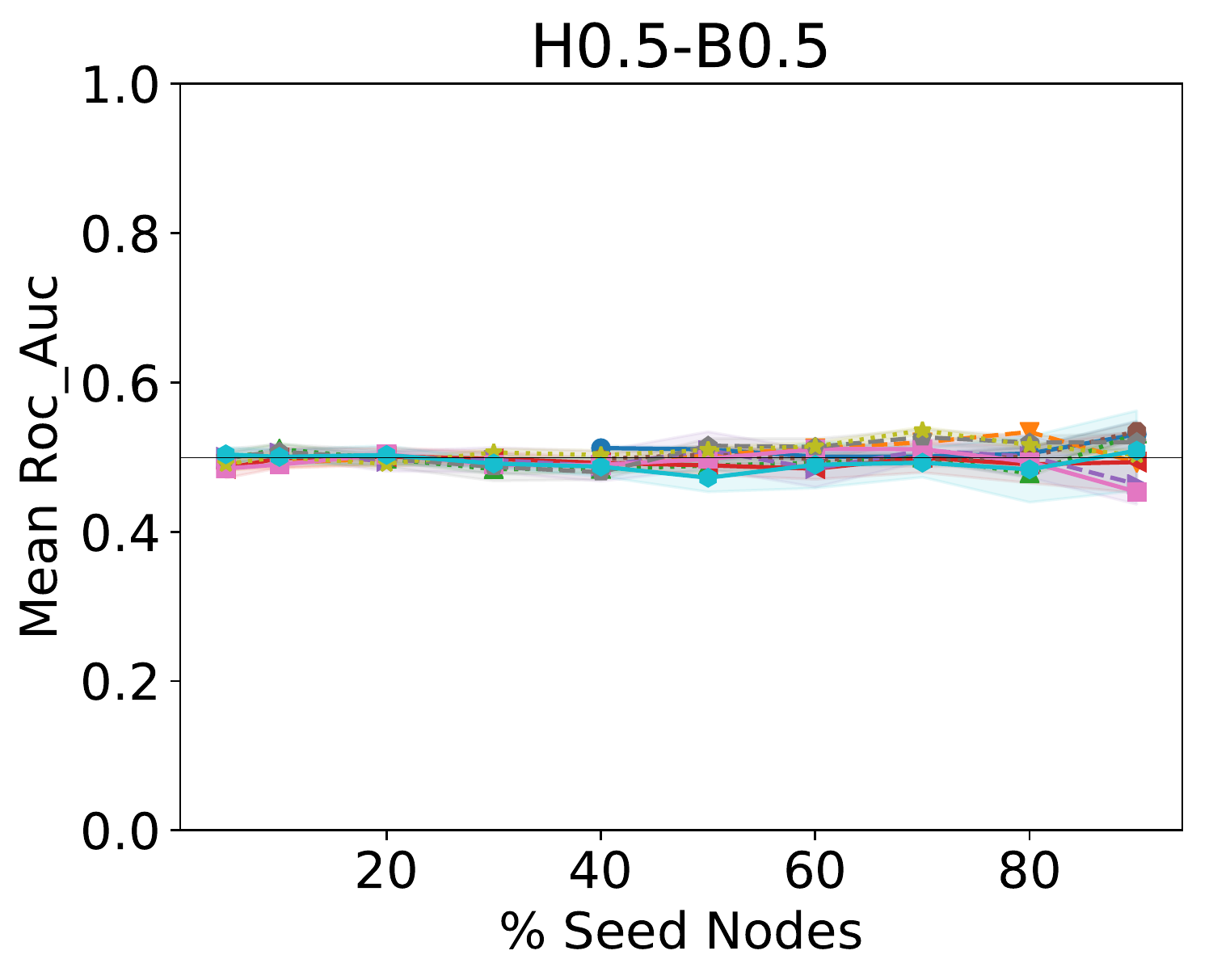}
        \caption{Neutral}
        \label{synthetic-rocauc:neutral}
    \end{subfigure}
    ~ 
    \begin{subfigure}[t]{0.32\textwidth}
        \centering
        \includegraphics[height=1.8in]{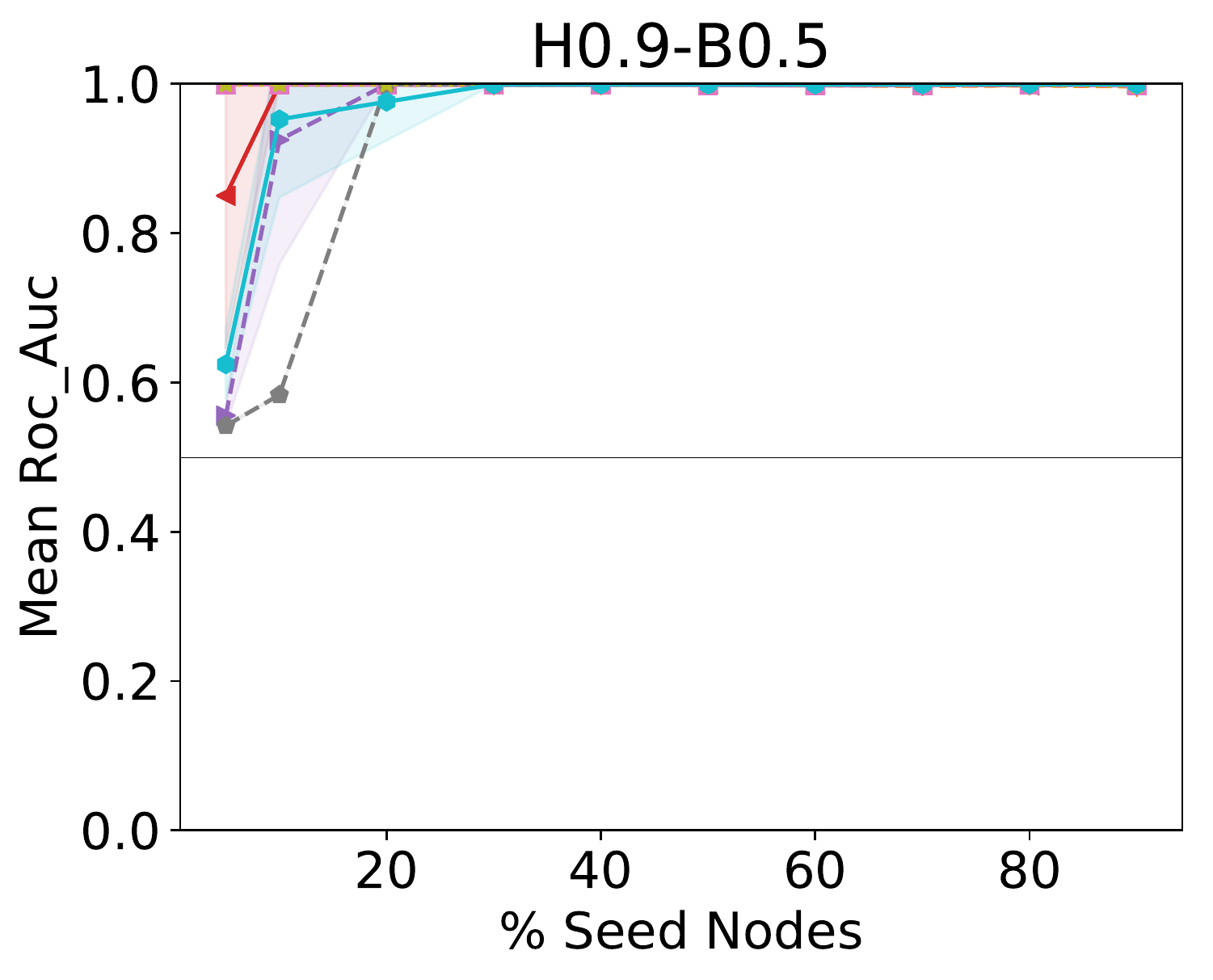}
        \caption{Homophilic}
        \label{synthetic-rocauc:homo}
    \end{subfigure}
    \caption{\textbf{Results on synthetic (sparse) networks ($\langle k \rangle=8$, $d=0.0039$)}. This figure shows the mean ROC-AUC values of classification for 10 sampling methods on (a) heterophilic, (b) neutral and (c) homophilic networks generated using the preferential attachment-based algorithm proposed by Karimi et.  al.  in \cite{karimi_homophily}. Sample size is shown on the x-axis. Values are averages of 5 runs; shaded areas depict standard deviations.}
    \label{synthetic-rocauc}
\end{figure*}

\section{Experiments}
\label{experiments}
The classification process can be summarised into three steps. First, it splits the nodes of the network into training and testing sets, then it learns the local and relational models using the subgraph extracted from the training sample, and finally it runs the classification on the testing set. The selection of nodes in the training sample varies depending on the sampling method. 
However, to compare sampling methods, the size of the training sample is kept constant and contains between $5-90\%$ of nodes from $V$.

\subsection{Synthetic Networks}
\label{experiments:synthetic}
\para{Datasets}. We generate $11$ networks with given class balance ${B=0.5}$, homophily ${H\in\{0.0,0.1, \dots , 1.0\}}$ and starting degree $m=4$, using the preferential attachment-based algorithm proposed by Karimi et. al. in~\cite{karimi_homophily}. Every network consists of $N=2000$ nodes, $|E|=7984$ edges, average degree $\langle k \rangle=8$, and link density ${d=0.0039}$\footnote{Since link density is very small, we refer to this set of networks as sparse networks.}. \Cref{tbl:network-properties} shows more network properties for some heterophilic, neutral and homphilic networks.
In general, each node is assigned one binary attribute, i.e., $color \in \{blue, red\}$, which defines its class membership. 
The probability of node $j$ to connect to node $i$ is given by:
\begin{equation}
\Pi_{i} = \frac{h_{ij} k_{i}}{\sum_{l} h_{lj} k_{l}}
\label{eq:homophilic_BA}
\end{equation}
where $k_{i}$ is the degree of node $i$ and $h_{ij}$ is the homophily between the two nodes \cite{karimi_homophily}. 
Considering that homophily is symmetric and complementary, we can assume that the probability of connections between nodes that belong to class blue ($h_{bb}$) and nodes that belong to class red ($h_{rr}$) is identical ($h_{bb} = h_{rr} = H$) and is complementary to intra-class link probability  $h_{br} = h_{rb} = 1 - H$.
We vary homophily from $H=0.0$ (completely heterophilic) to $H=1.0$ (completely homophilic). When $H=0.0$, only nodes that do not share the same attribute are connected.
In contrast, in the complete homophilic case, only nodes that share the same attribute are interlinked. In neutral networks ($H=0.5$), a node is equally likely to link to nodes with either label. That means, in neutral networks the formation of edges is statistically independent from node attributes. 

\begin{table}[!b]
\small
\centering
\caption{\textbf{Synthetic (sparse) network properties}. This table shows properties of the networks analysed in this work. These networks contain two balanced groups of nodes (i.e., blue, red). Each numeric column represents a single network with a specific level of homophily.}
\label{tbl:network-properties}
\begin{tabular}{@{}lrrr@{}}
\toprule
\multicolumn{1}{c}{\multirow{2}{*}{\textbf{Property}}} & \multicolumn{3}{c}{\textbf{Homophily}} \\ \cmidrule(l){2-4} 
\multicolumn{1}{c}{} & \multicolumn{1}{r}{\textbf{0.1}} & \multicolumn{1}{r}{\textbf{0.5}} & \multicolumn{1}{r}{\textbf{0.9}} \\ \midrule
{Attribute Assortativity} & -0.8 & 0.01 & 0.8 \\
{Clustering Coefficient} & 0.01 & 0.02 & 0.03 \\
{Link Density} & 0.0039 & 0.0039 & 0.0039 \\
{Degree Assortativity} & -0.06 & -0.06 & -0.05 \\
{Node Connectivity} & 4 & 4 & 4 \\ \bottomrule
\end{tabular}
\end{table}

\para{Results}.
For simplicity, we report on networks with special cases of homophily, i.e., ${H\in\{0.1,0.5,0.9\}}$.
These results are shown in \Cref{synthetic-rocauc}. 
From \Cref{synthetic-rocauc:neutral} we see that classification performance across all sampling methods is uniform in neutral networks since the formation of links is independent of the node attributes. Therefore, relational classifiers cannot detect any pattern in the network structure that helps to guess the correct attributes. 
The comparison between heterophilic (\Cref{synthetic-rocauc:hete}) and homophilic (\Cref{synthetic-rocauc:homo}) networks shows that regardless of the sampling technique and sample size, heterophilic networks are easier to classify (i.e., all ROC-AUC values are $1.0$), whereas homophilic networks (in some cases) require larger training samples to achieve perfect classification.
For instance, the overall classification performance is worse (ROC-AUC $\approx$ 0.6) for sampling by nodes, percolationASC, snowball and nedges, if sample sizes are very small ($5\%$). However, once sample sizes increase, ROC-AUC values quickly converge to $1.0$.

\begin{figure*}[hpt]
    \centering
    \includegraphics[height=0.45in]{{legend-h}.pdf}
    
    \centering
    \begin{subfigure}[t]{0.32\textwidth}
        \centering
        \includegraphics[height=1.8in]{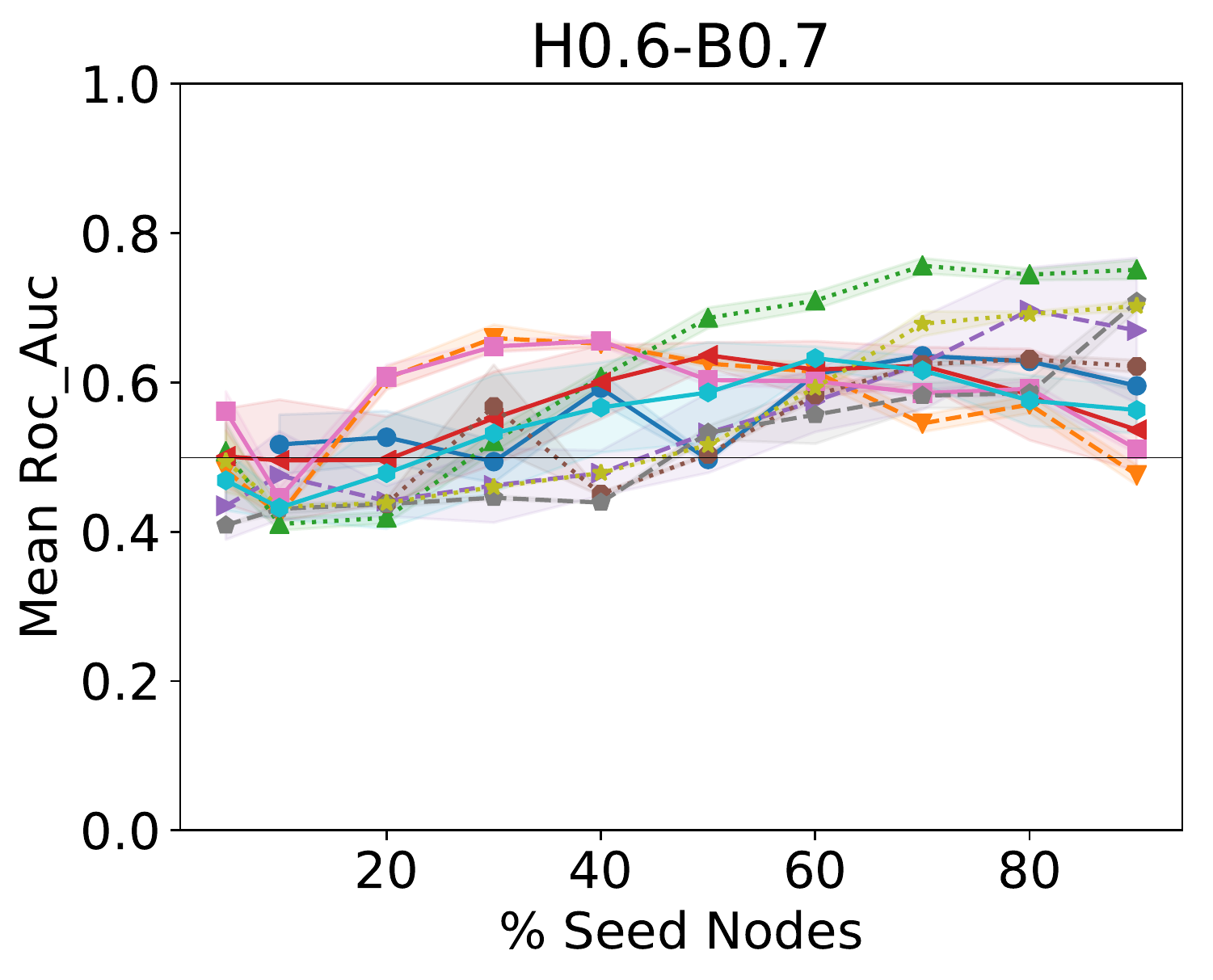}
        \caption{ROC-AUC}
        \label{caltech:rocauc}
    \end{subfigure}
    ~ 
    \begin{subfigure}[t]{0.32\textwidth}
        \centering
        \includegraphics[height=1.8in]{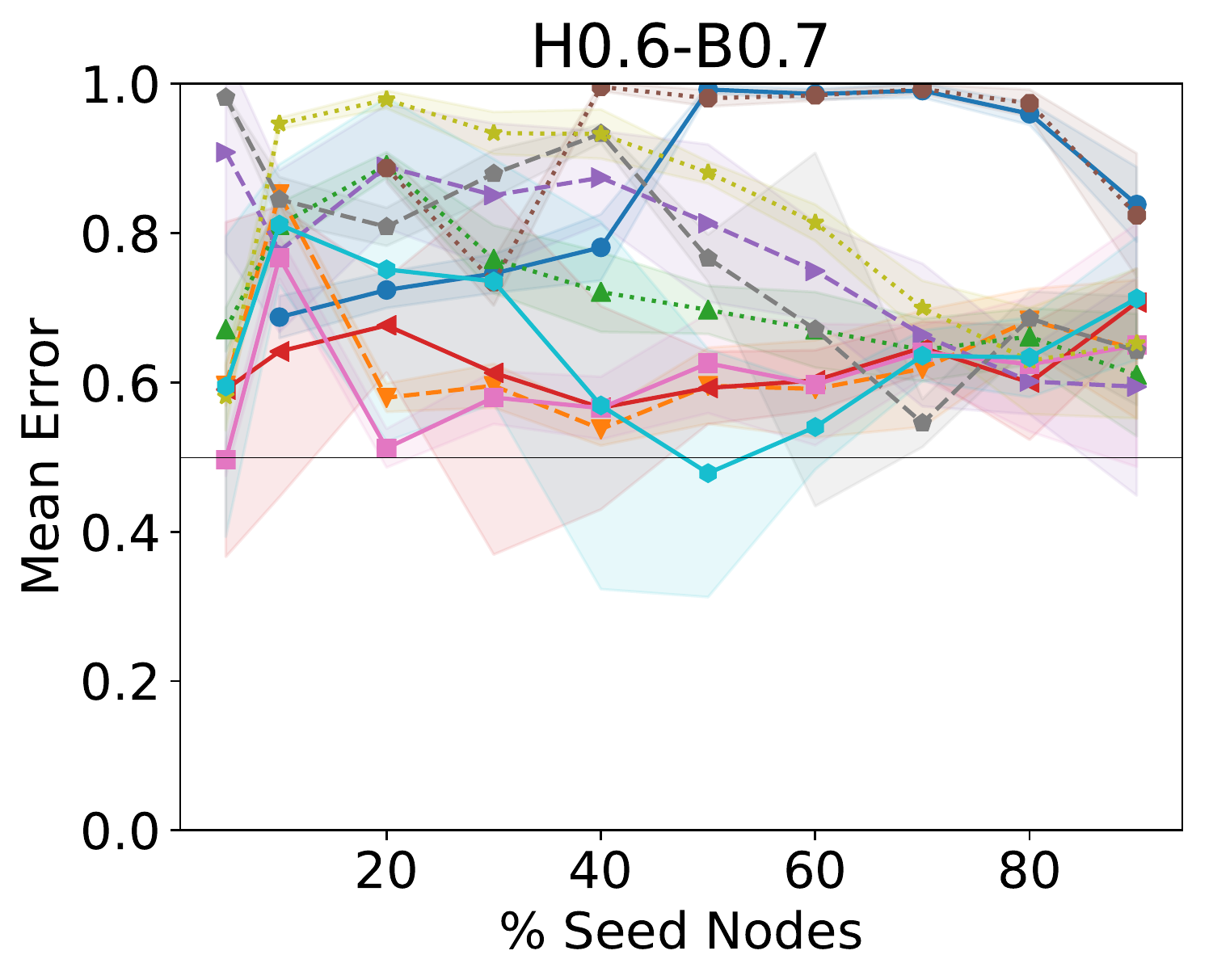}
        \caption{Mean Error gender 1 (minority)}
        \label{caltech:minority}
    \end{subfigure}
    ~ 
    \begin{subfigure}[t]{0.32\textwidth}
        \centering
        \includegraphics[height=1.8in]{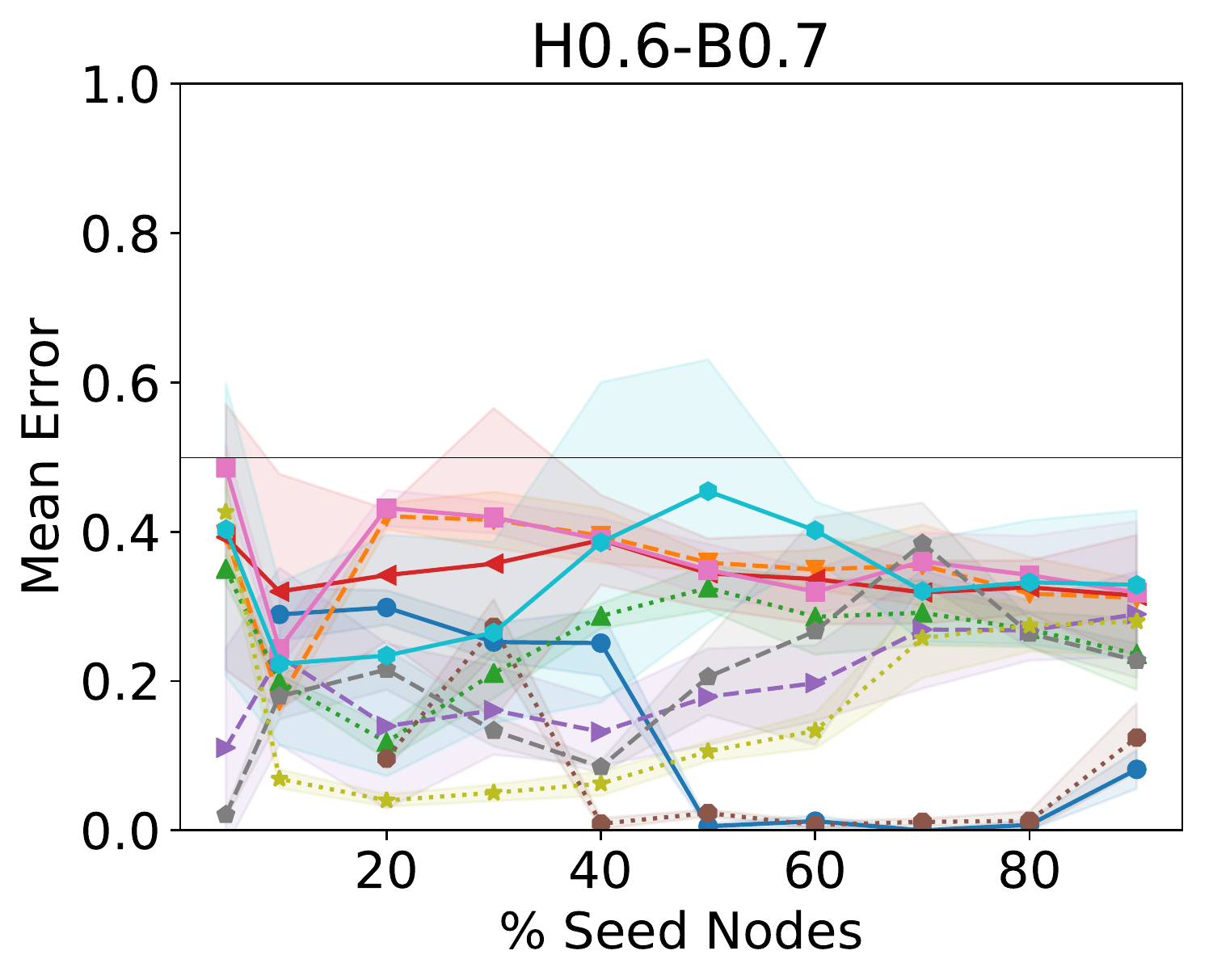}
        \caption{Mean Error gender 2 (majority)}
        \label{caltech:majority}
    \end{subfigure}
    \caption{\textbf{Results on Caltech Facebook dataset}. From left to right, this figure shows the performance of 10 different sampling techniques on the CalTech dataset for different sample sizes: (a) mean ROC-AUC, (b) classification mean error of class gender~1, and (c) classification mean error of class gender~2. Values are averages of 5 runs; shaded areas depict standard deviations. Recall, this network is unbalanced $B=0.7$ towards gender~2, and somewhat neutral $H=0.6$. Thus, it is not surprising that classification performance is around $ROC-AUC=0.5$. From (a) we can see that a small training sample of $30\%$ of highest degree nodes (degreeDESC) can achieve $ROC-AUC=0.66$. However, the best model is degreeMIX, which achieves $ROC-AUC=0.76$ with $70\%$ of top mix degree nodes (i.e., $35\%$ top high degree, and $35\%$ top low degree nodes). In general, all sampling methods improve the classification performance with higher sample sizes. From figures (b,c) we can see the class imbalance problem. Since gender~2 is the majority class, it has lower classification error compared to the minority class gender~1.}
    \label{caltech}
\end{figure*}

This asymmetry between homophilic and heterophilic networks is clear in \Cref{fig-summary:sparse-hete,fig-summary:sparse-homo}, which summarises the classification performance on samples drawn from balanced networks using different sampling strategies. Color represents different sampling methods, and bars the minimum sample size required so that a classifier 
achieves an error below $20\%$ for both classes.
Homophilic networks require larger sample sizes to achieve good classification performance for random nodes, nedges and snowball sampling, as well as all metrics ranked in ascendant order (generally, lowest degree nodes first). Hence, sampling methods that are biased towards higher degree nodes (DESC) outperform the other techniques. Heterophilic networks on the other hand are easier to classify, since 8 out of 10 sampling methods are good enough to achieve good performance for both classes with small training samples that contain only $5\%$ of the nodes. Notice that for the particular cases of degreeASC and pagerankASC, all networks require at least $40\%$ of seed nodes to achieve good performance. This occurs because both sampling techniques rank lowest-degree nodes first, and these nodes do not necessary link to each other\footnote{In fact, degree assortativity\cite{barabasi2016network} for these networks is around $-0.06$ (i.e., no degree correlation).}.
Hence, such samples contain disconnected nodes (i.e., $\hat{E}=\varnothing$), not enough for learning the model's parameters.

\subsection{Real-World Networks}
\label{realworld}

\para{Dataset}.
We choose one of the 100 Facebook networks extracted back in 2005~\cite{traud2012social}. We focus on the Caltech University network which includes only intra-school links (i.e., friendship links between user's FB pages). Every node represents a member of the school, and it is described by several attributes: a student/faculty status
flag, gender, major, second major/minor, dorm/house,
year, and high school. For the purpose of our experiments we choose the attribute $gender \in \{1,2\}$ as the class label (and only attribute) for the classifier. After removing nodes without gender information (i.e., $gender=0$), and nodes with no edges we end up with $701$ nodes and 15464 edges. The final network is somewhat neutral ($H=0.6$), and unbalanced ($B=0.7$) towards gender~2. Properties of this network are shown in~\Cref{caltech:properties}. For instance, we can see that people are highly connected (i.e., $44.12$ friendships on average). 

Notice that this network differs from the synthetic network examples, regarding not only class imbalance, but also average degree, link density and clustering coefficient.

Although this network goes beyond the scope of our work (i.e., it is an unbalanced network), we include it in this report for two reasons: (i) to highlight the importance of further research on minimising the class imbalance problem, and (ii) to show whether homophily has an impact on classification regardless of class imbalance.

\begin{table}[!hb]
\centering
\caption{\textbf{CalTech 2005}. Properties of the CalTech university Facebook network.}
\label{caltech:properties}
\begin{tabular}{@{}lllll@{}}
\cmidrule(r){1-2} \cmidrule(l){4-5}
\multicolumn{1}{l}{\textbf{Property}} & \multicolumn{1}{l}{\textbf{Value}} &  & \multicolumn{1}{l}{\textbf{Property}} & \multicolumn{1}{l}{\textbf{Value}} \\ \cmidrule(r){1-2} \cmidrule(l){4-5} 
N & 701 &  & $\langle k \rangle$ & 44.12 \\
|E| & 15464 &  & $\langle k_{minority} \rangle$ & 51 \\
gender~1 (min.) & 228 (33\%) &  & $\langle k_{majority} \rangle$ & 41 \\
gender~2 (maj.) & 473 (67\%) &  & node connectivity & 0 \\B & $\sim$0.70 &  & degree assortativity & -0.0617 \\
H & 0.6 &  & attribute assortativity & 0.054 \\
density & 0.063 &  & clustering & 0.39 \\  \cmidrule(r){1-2} \cmidrule(l){4-5}
\end{tabular}
\end{table}

\begin{figure*}[tbh!]
\centering
    \includegraphics[height=0.45in]{{legend-h}.pdf}
    
    \centering
    \begin{subfigure}[t]{0.32\textwidth}
        \centering
        \includegraphics[height=1.8in]{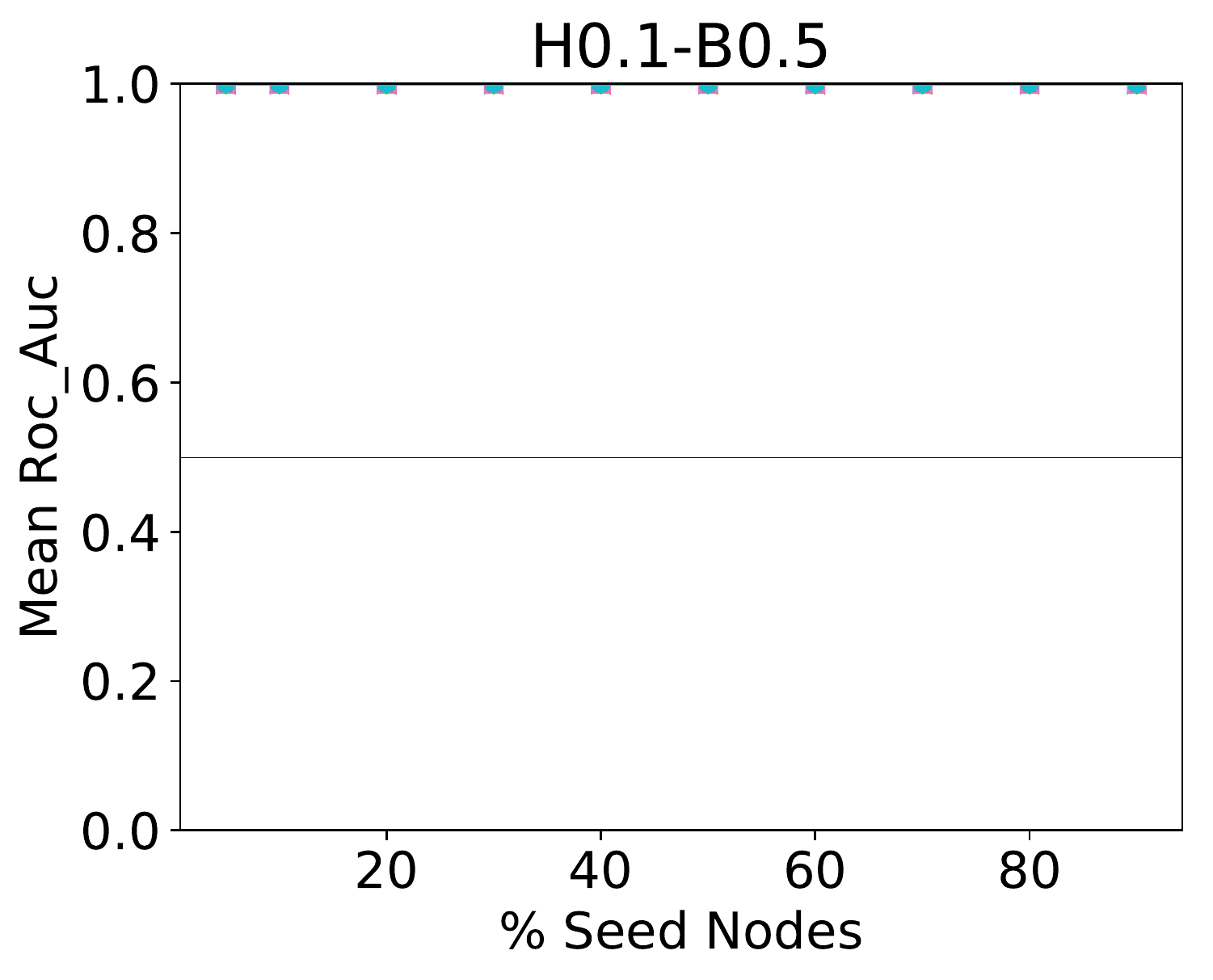}
        \caption{Heterophilic}
        \label{synthetic-rocauc-dense:hete}
    \end{subfigure}
    ~ 
    \begin{subfigure}[t]{0.32\textwidth}
        \centering
        \includegraphics[height=1.8in]{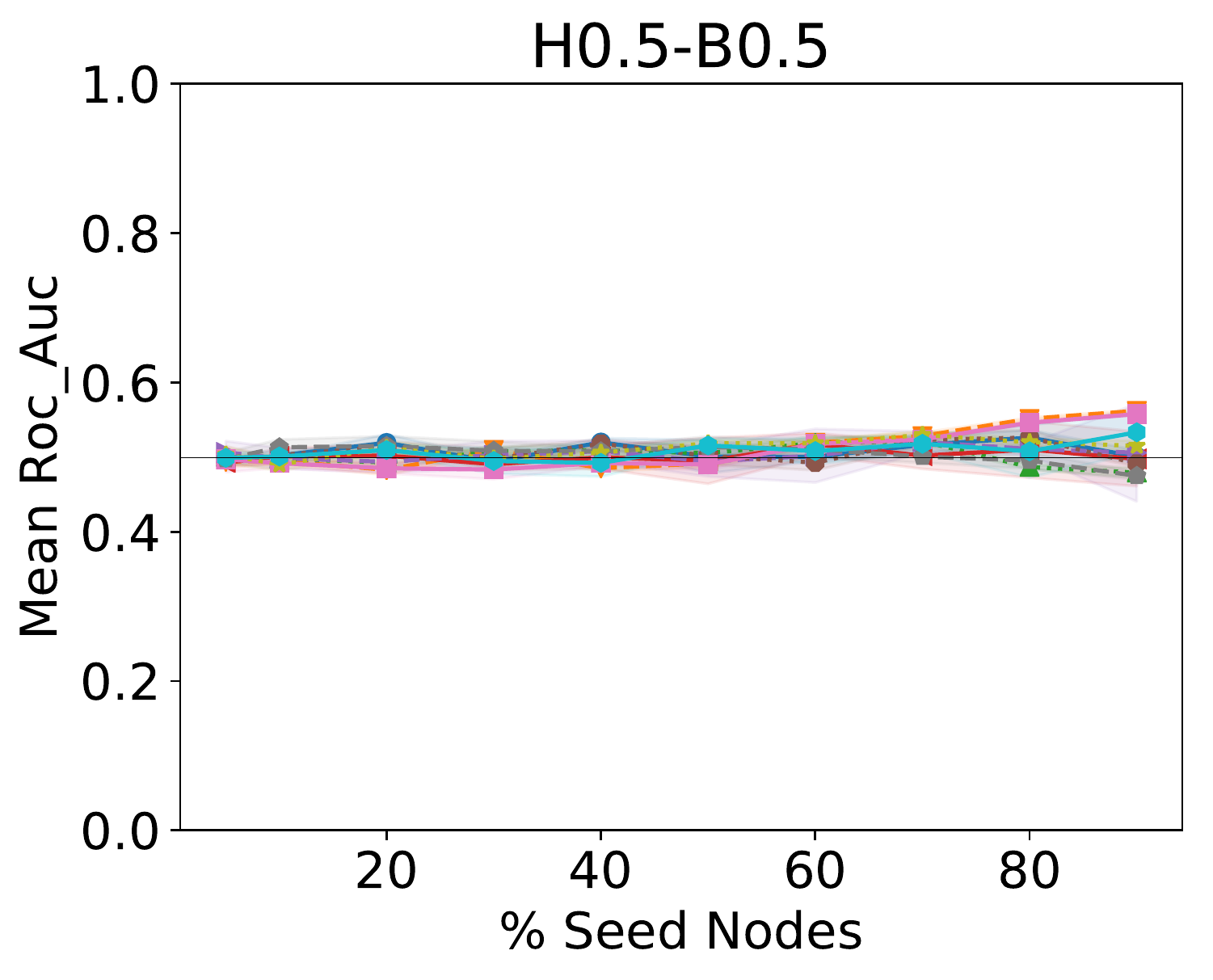}
        \caption{Neutral}
        \label{synthetic-rocauc-dense:neutral}
    \end{subfigure}
    ~ 
    \begin{subfigure}[t]{0.32\textwidth}
        \centering
        \includegraphics[height=1.8in]{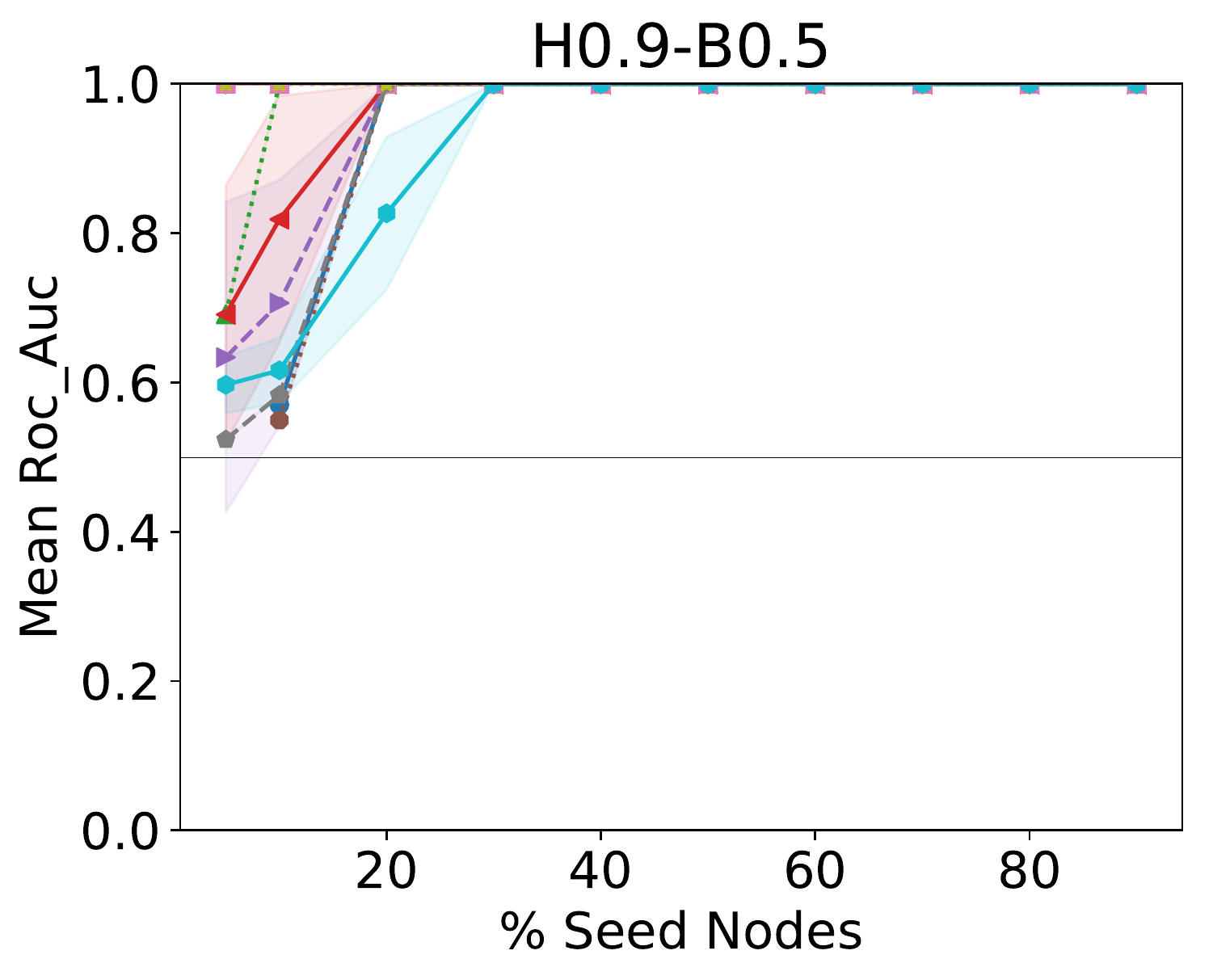}
        \caption{Homophilic}
        \label{synthetic-rocauc-dense:homo}
    \end{subfigure}
    \caption{\textbf{Results on synthetic (dense) networks ($\langle k \rangle=40$, $d=0.019$)}. Similar to \Cref{synthetic-rocauc}, this figure shows the mean ROC-AUC values of classification for 10 sampling methods on (a) heterophilic, (b) neutral and (c) homophilic networks. The difference relies on link density. These networks posses higher density. Sample size is shown on the x-axis. Values are averages of 5 runs; shaded areas depict standard deviations.}
    \label{synthetic-rocauc-dense}
\end{figure*}

\para{Results}.
\Cref{caltech} shows the classification results of the CalTech network. Since this network is somewhat neutral ($H=0.6$) we expect its performance to be similar to a uniform classifier (i.e., random guessing). 
\Cref{caltech:rocauc} confirms this expectation, since it shows that most sample techniques achieve a ROC-AUC value around $0.5$, especially for small samples ($5-50\%$ seed nodes). We conclude that degreeMIX outperforms the other sampling methods, although it requires at least 70\% of the total number of nodes $N$ to achieve a ROC-AUC=0.76.

 ROC-AUC values give us an overall performance of the classification, but they do not give us the whole picture within classes. In \Cref{caltech:minority,caltech:majority}, the classification mean error values for minority and majority are shown respectively. Here we observe the \emph{class imbalance} problem, where classification estimates tend to lean towards the majority class: mean error values for gender~2 (i.e., majority $70\%$) are lower than for gender~1 (i.e., minority $30\%$).

These results show the importance of further research on the understanding of relational classification in unbalanced networks with different levels of homophily.

\section{Discussion}
\label{discussion}
The current work presents a descriptive study of the effect of network sampling on the performance of relational classification. In the following we present a detailed discussion of the factors we explored.

\subsection{Network Structure}

\para{Link Density}.
The synthetic networks used in our experiments were generated with $N=2000$ nodes, $|E|=7984$ edges, average degree $\langle k \rangle=8$, and link density $d=0.0039$. Previous work found that link density impacts performance of relational classifiers \cite{sen08,zenoinvestigating}. To test this finding, we increased link density to $d=0.019$, which resulted in ${|E_{dense}|=39600}$ edges, and average degree ${ \langle k_{dense} \rangle=40}$. We refer to this set of networks as dense networks. As we see in \Cref{synthetic-rocauc-dense}, higher link density did not improve the classification performance.
For neutral (\Cref{synthetic-rocauc-dense:neutral}) and heterophilic (\Cref{synthetic-rocauc-dense:hete}) networks, results were similar to the ones using networks with lower link density. Classification of homophilic networks (\Cref{synthetic-rocauc-dense:homo}) on the other hand, worsened ROC-AUC values for very small sample sizes (i.e., 5-20\% seed nodes). Only degreeDESC, pagerankDESC and percolationDESC outperformed the other techniques while using only 5\% of nodes. Notice that in both variants of link density, ROC-AUC values converged to 1.0 for all sampling strategies with at least 30\% of seed nodes.

\para{Class Imbalance}.
Real-world networks can be highly unbalanced, with dissimilar proportions of nodes of each type. For example, the network we studied in \Cref{realworld} has homophily $H=0.6$, and class balance $B=0.7$. We showed that class balance affects classification results. Due to the almost neutral homophily, the collective inference performed only slightly better than random baseline 
(i.e., random guessing), regardless of the sampling technique. However, due to class imbalance the error for each class (e.g., minority vs. majority) differed drastically. For instance, using random node sampling and 30\% seed nodes, class gender 1 achieved a classification error of 0.85, whereas class gender 2 only 0.16.
Further research is needed to understand  dynamics of relational classification in unbalanced networks.

\begin{figure*}[tbh!]

    \centering
    \begin{subfigure}[t]{0.32\textwidth}
        \centering
        \includegraphics[height=1.8in]{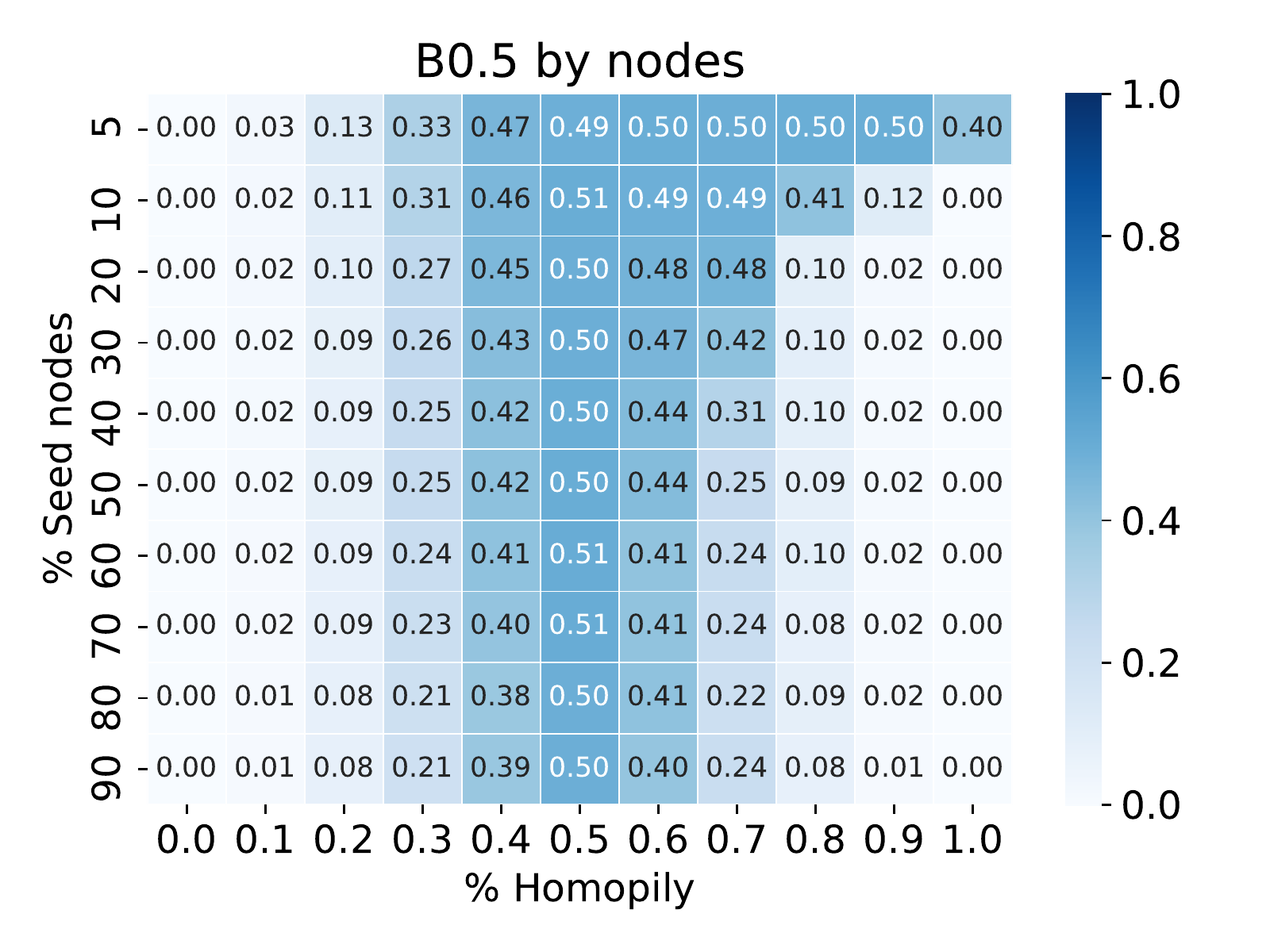}
        \caption{nodes Sampling}
        \label{fig-synthetic-error-balance:nodes}
    \end{subfigure}
    ~ 
    \centering
    \begin{subfigure}[t]{0.32\textwidth}
        \centering
        \includegraphics[height=1.8in]{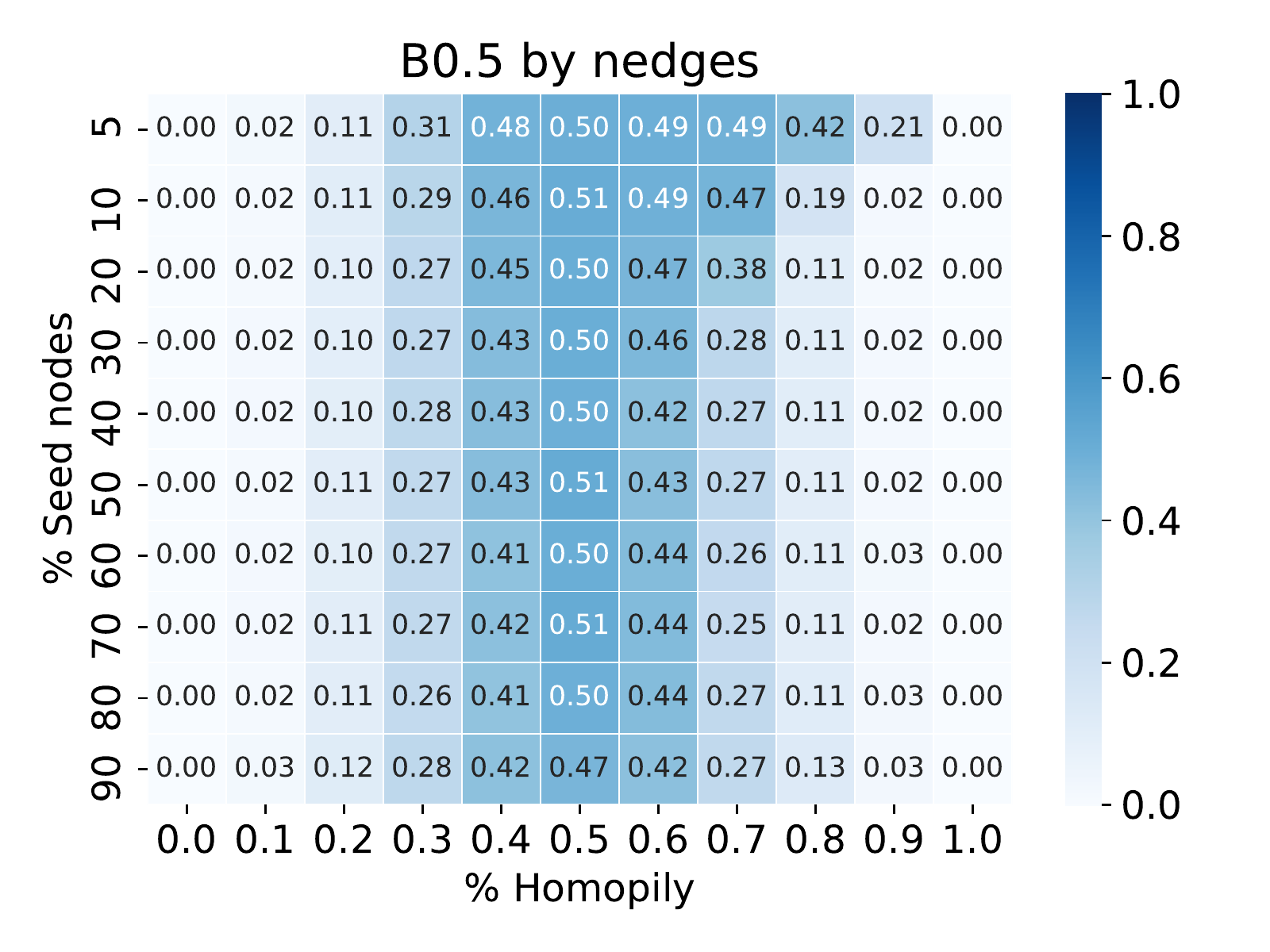}
        \caption{nedges Sampling}
        \label{fig-synthetic-error-balance:nedges}
    \end{subfigure}
    ~
    \centering
    \begin{subfigure}[t]{0.32\textwidth}
        \centering
        \includegraphics[height=1.8in]{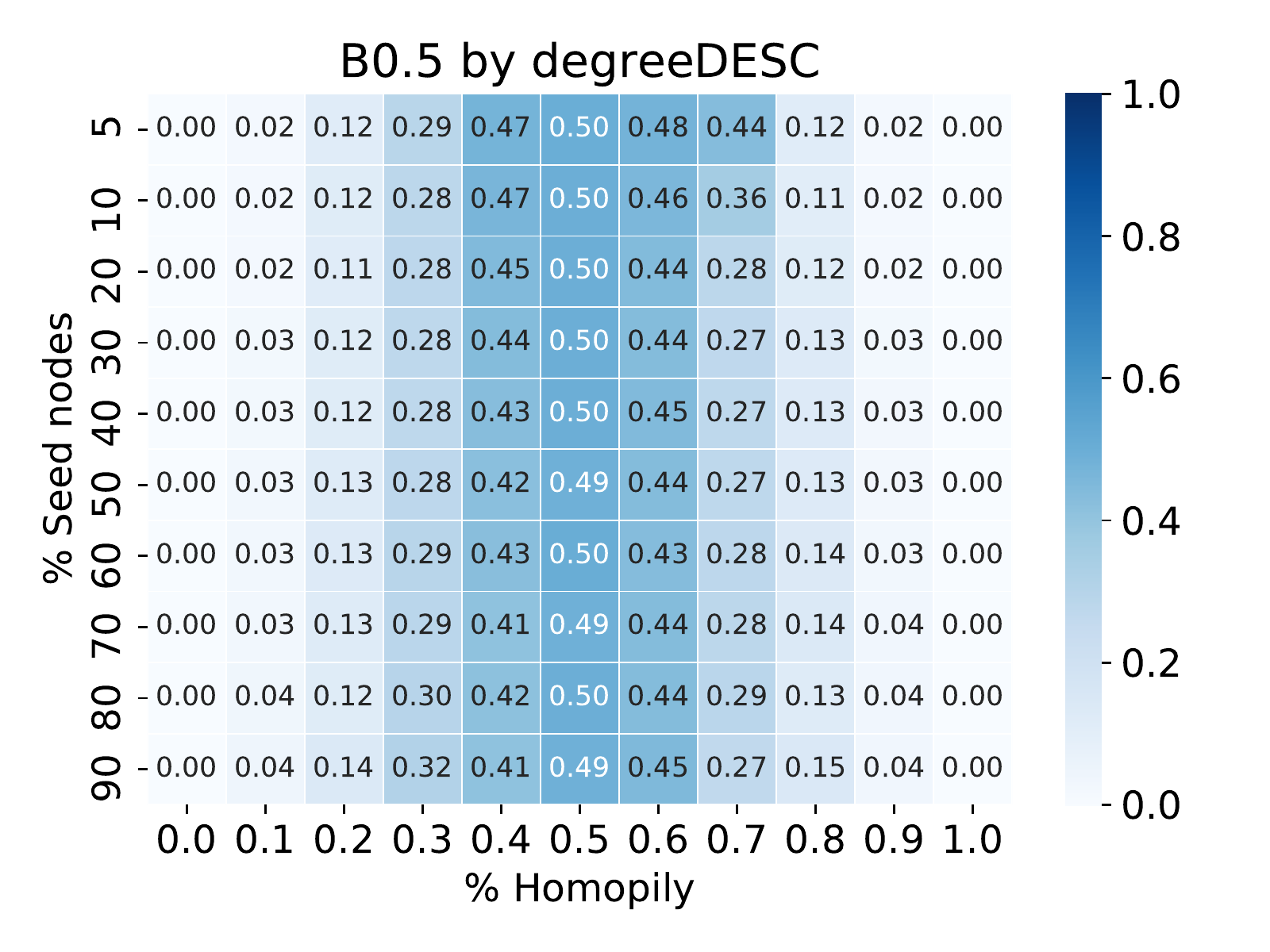}
        \caption{degreeDESC Sampling}
        \label{fig-synthetic-error-balance:degreeDESC}
    \end{subfigure}
    
  \caption{\textbf{Overall mean error of synthetic (sparse) networks}. These heatmaps illustrate the overall classification mean error using (a) random node sampling, (b) random edge sampling, and (c) degree descendant sampling. Columns represent networks of different homophily, from heterophilic ($H=0.0$) to homophilic ($H=1.0$). Every row shows the percentage of nodes collected in the sampling. Somewhat neutral networks (${0.4\leq H \leq0.6}$) perform uniformly using either sampling technique. 
  In general, the more heterophilic the network the more accurate the classification in all cases. However, classification error reduces with larger training samples for both node and edge sampling, whereas degreeDESC sampling works best using only $5\%$ of seed nodes.
  Homophilic networks on the other hand, require larger samples in order to achieve good performance using random node or edge sampling. Values are averages of 5 runs.}
  \label{fig-synthetic-error-balance}
\end{figure*}

\para{Homophily}.
Our work shows that homophily clearly impacts the performance of relational classifiers (cf. \Cref{synthetic-rocauc}). When networks are balanced, ROC-AUC curves vary depending on the level of homophily and sample size. As expected, in neutral networks (homophily $H=0.5$) all sampling methods perform equally well and sample size does not impact classification performance. Thus, it is not surprising that the classifier cannot learn any pattern, since no pattern exists, no matter the size of the sample.
However, in heterophilic  ($H=0.1$) or homophilic ($H=0.9$) networks, the classification accuracy varies based on the sampling strategy and number of labelled seeds present in the training sample. 
To understand this, let us focus on the performance of three different sampling methods over networks with different levels of homophily shown in \Cref{fig-synthetic-error-balance}. Each heatmap shows the classification mean error (averaged over 5 runs) for each of the 11 synthetic networks (x-axis) described in \Cref{experiments:synthetic}, and the amount of seed nodes in the training sample (y-axis). Darker cells represent higher errors, i.e., worse performance.

Overall, we can see that in the heterophilic regime (H $\leq$ 0.2, leftmost columns), the classifier works very well with a small fraction of seed nodes. Moreover, classification error reduces with larger training samples for node and edge sampling, and with smaller training samples for degreeDESC sampling.
For homophilic networks (H $\geq$ 0.8, rightmost columns), the classification error is high when the training sample is small for node and nedge sampling. DegreeDESC on the other hand, performs best with only 5\% of seed nodes.
This seems counter-intuitive, as one would expect perfect classification in both cases, since the strong homophily and strong heterophily should help the classifier learn the relationships between links and attributes. Moreover, global properties of both networks seem to be almost identical, as shown in \Cref{tbl:network-properties}.
At first glance, it seems that heterophilic networks are easier to classify and their performance do not vary across sampling techniques. Furthermore, high degree nodes help the classifier to not only learn the correct parameters, but also to propagate the correct inference among nodes in both heterophilic and homophilic networks.

\subsection{Sampling Networks}
\begin{figure*}[!ht]
    \centering
    \includegraphics[height=0.45in]{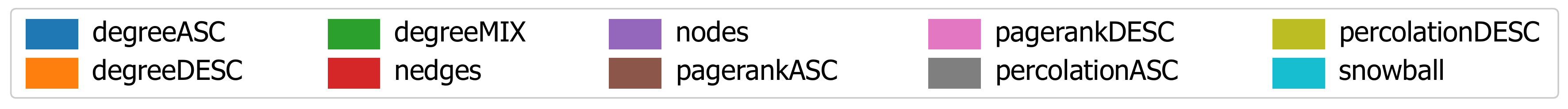}
    
    \centering
    \begin{subfigure}[t]{0.45\textwidth}
        \centering
        \includegraphics[height=1.5in]{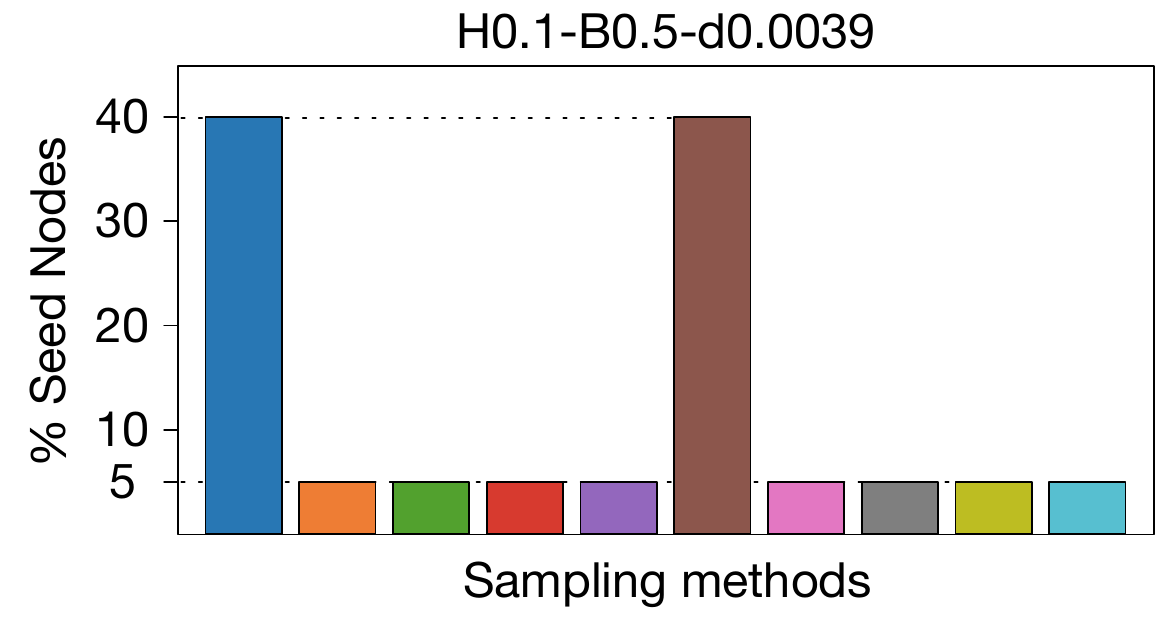}
        \caption{Sparse and Heterophilic}
        \label{fig-summary:sparse-hete}
    \end{subfigure}
    ~ 
    \begin{subfigure}[t]{0.45\textwidth}
        \centering
        \includegraphics[height=1.5in]{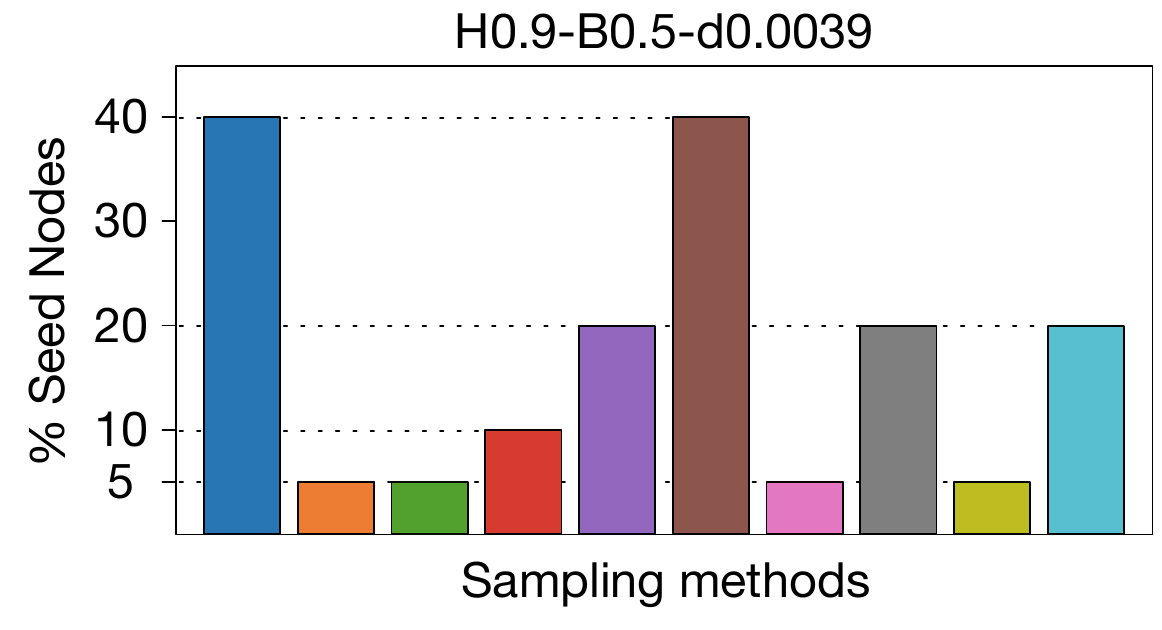}
        \caption{Sparse and Homophilic}
        \label{fig-summary:sparse-homo}
    \end{subfigure}
    
    \begin{subfigure}[t]{0.45\textwidth}
        \centering
        \includegraphics[height=1.5in]{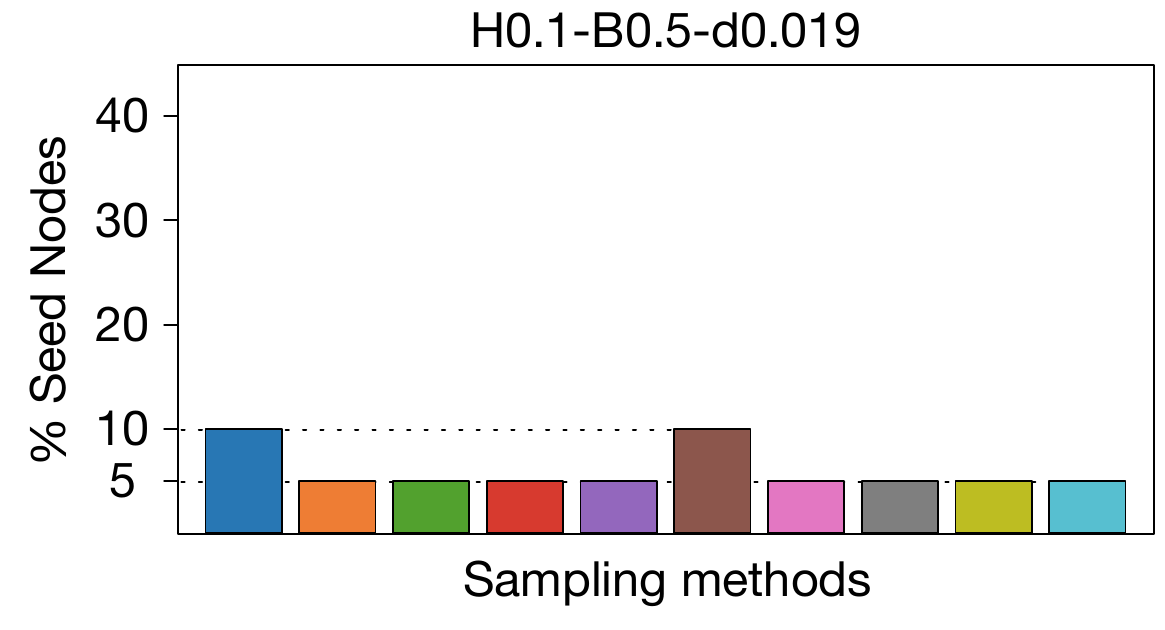}
        \caption{Dense and Heterophilic}
        \label{fig-summary:dense-hete}
    \end{subfigure}
    ~ 
    \begin{subfigure}[t]{0.45\textwidth}
        \centering
        \includegraphics[height=1.5in]{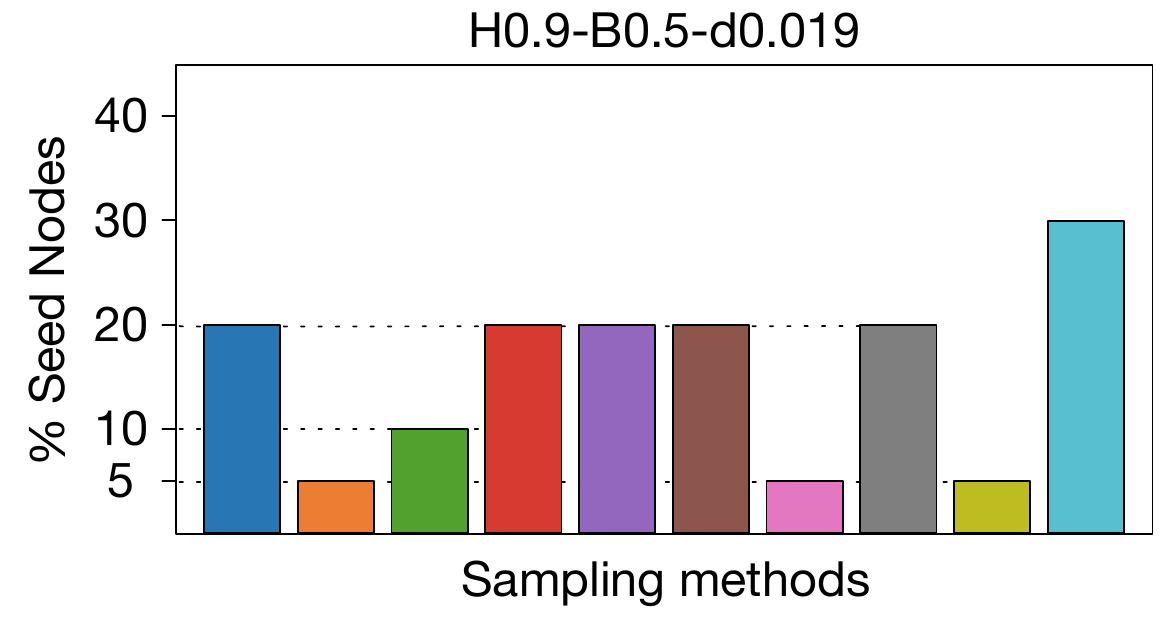}
        \caption{Dense and Homophilic}
        \label{fig-summary:dense-homo}
    \end{subfigure}
    \caption{\textbf{Ranking of Sampling methods}. This figure depicts the minimum sample size (in terms of percentage of nodes) required for all sampling techniques in order to achieve a classification error below $20\%$ for both classes (blue and red) in balanced scale-free networks. We summarise from left to right, (a,c) heterophilic and (b,d) homophilic networks, and from top to bottom, (a,b) sparse and (c,d) dense networks. Each color represents a different sampling method. The lower the bars the better. For instance, sampling by degreeDESC (second bar from left to right in each plot) works very well by picking the top5\% of highest degree nodes in all cases. In general, classification works best for heterophilic networks since they require very small training samples regardless of sampling strategy and density.}
    \label{fig-summary}
\end{figure*}

As described in \Cref{section:sampling}, we used  ten different network sampling strategies. 
In \Cref{fig-summary} we rank sampling methods based on the sample size that is required to train a classifier that achieves a classification error below $20\%$ for both classes\footnote{While this measure might be arbitrary, the goal is to show an unbiased classification, where both classes get correctly inferred}. \Cref{fig-summary:sparse-hete,fig-summary:sparse-homo} refer to the sparse networks shown in \Cref{{experiments:synthetic}} (link density $d=0.0039$). \Cref{fig-summary:dense-hete,fig-summary:dense-homo} refer to the respective dense versions (link density $d=0.019$).
The rightmost figures show the results for classifiers trained on homophilic networks. One can see that sampling strategies that include nodes which have a central position in the network (degreeDESC, pagerankDESC, percolationDESC) work best, and require only $5\%$ of seed nodes. However, these sampling strategies require knowledge of the full network and may not be appropriate in cases where this information is difficult to obtain.
In those cases, the second best option for sparse networks (cf. \Cref{fig-summary:sparse-homo}) is to sample $10\%$ of nodes by nedges, which randomly selects edges from the network. Notice that this leads to hubs (i.e., high degree nodes) being preferred. For dense networks (cf. \Cref{fig-summary:dense-homo}), the second best option is to sample 20\% of random nodes.

The rest of sampling techniques require larger training samples. Next, we explain their poor performance. Snowball sampling is a technique that randomly picks a node, then its neighbours, and the neighbours' neighbours, similar to breadth-first search. It is expected to require a larger sample of seed nodes, since in a homophilic network, a snowball sample that starts with a red node, will then select its neighbours---who are likely to be mostly red---without capturing enough blue nodes. 
Degree, pagerank and percolation ascendant sampling methods (degreeASC, pagerankASC, percolationASC) are usually ranked the worst. This is not surprising, since low degree and low pagerank nodes tend to connect to fewer nodes, leaving a vast majority of nodes disconnected from them (especially in sparse networks).

In the heterophilic regime (leftmost figures), degreeASC and pagerankASC are also ranked last. This is also due to the fact that low degree nodes connect to only a few nodes. However, eight out of ten sampling techniques require a training sample containing only 5\% of seed nodes. In other words, in a heterophilic network, it is enough to collect only 5\% of random nodes in order to achieve good classification performance\footnote{This holds for $0.0\leq H \leq 0.2$ as shown in \Cref{fig-synthetic-error-balance:nodes}.}.

\section{Conclusions and Future Work}
\label{conclusions}

In this paper we presented a step towards quantifying sampling bias in network inference. Precisely, we studied the influence of network structure and sampling on relational classification. Our findings are as follows. (i) Heterophilic networks are easier to classify than homophilic networks. The former require only 5\% of seed nodes to be able to classify correctly most unlabelled nodes. This holds for 8 out of 10 sampling methods. (ii) Link density and degree assortativity influence the performance of sampling methods that rank low-degree nodes first. Low-degree nodes are less likely to connect to each other if the network has low link density and negative or neutral degree assortativity. Therefore, sampling methods that rank low degree nodes first require larger samples to include not only more nodes, but also more edges. (iii) Link density also influences the performance of homophilic networks. The higher the link density, the larger the training samples for degreeMIX, nedges and snowball sampling in order to achieve good accuracy for all classes. 
(iii) High degree nodes are the best seed nodes, since only 5\% of them achieve optimal classification performance (ROC-AUC=1.0) for homophilic, heterophilic, dense and sparse networks. However, these sampling strategies require knowledge of the full network and may not be appropriate in cases where this information is difficult to obtain. Therefore, in those cases it is sufficient to sample: 5\% of random nodes in heterophilic networks, 10\% of nedges if the network is homophilic and has low link density, and 20\% nodes if the network is homophilic and has high link density.

How to select a sample that reflects the global properties of the original network and allows accurate label propagation is still an open research question. In future work, we plan to investigate the trade-off between constructing well-connected samples that help the classifier to learn the pattern between link formation and attributes and sampling nodes that are as distant as possible to gain information about different parts of the original network.
We also plan to investigate: (i) networks with multiple attributes where attribute distributions are skewed (i.e., imbalanced classes exist), (ii) more sophisticated sampling techniques such as the one in~\cite{avrachenkov2016inference}, and (iii) other relational models such as relational logistic regression\cite{yang2017stochastic}.

\para{Acknowledgements}.
The authors would like to thank Dr. Peter Fennell for his time and helpful advice on various technical topics, and Prof. Dr. Markus Strohmaier, Dr. Mathieu G\'enois and Reinhard Munz for their valuable comments and suggestions to improve the quality of the paper. This work was supported in part by the ARO (contract W911NF-16-1-0306) and by the AFOSR (contract FA9550-17-1-0327).

\bibliographystyle{ACM-Reference-Format}

\end{document}